\documentclass[10pt,twocolumn]{article}

\usepackage{graphicx}
\usepackage{subcaption}
\usepackage{amsmath,amssymb,amsfonts}
\usepackage[bookmarks=false]{hyperref}
\usepackage{url}
\usepackage[gen]{eurosym}
\usepackage{authblk}

\begin{document}
\title{Novel market approach for locally balancing renewable energy production and flexible demand}
\author[*]{Jos\'e Horta}
\author[*]{Daniel Kofman}
\author[**]{David Menga}
\author[***]{Alonso Silva}
\affil[*]{Telecom Paristech, 23 Avenue d'Italie, Paris, France\\ Emails: \{jose.horta, daniel.kofman\}@telecom-paristech.fr}
\affil[**]{EDF R\&D, EDF Lab Paris-Saclay, 91120 Palaiseau, France\\ Email: david.menga@edf.fr}
\affil[***]{Nokia Bell Labs, Nokia Paris-Saclay, Route de Villejust, 91620 Nozay, France\\ Email: alonso.silva@nokia-bell-labs.com}
\date{}

\maketitle

\begin{abstract}
Future electricity distribution grids will host a considerable share of variable renewable energy sources and local storage resources. Moreover, they will face new load structures due for example to the growth of the electric vehicle market. These trends raise the need for new paradigms for distribution grids operation, in which Distribution System Operators will increasingly rely on demand side flexibility and households will progressively become prosumers playing an active role on smart grid energy management. However, in present energy management architectures, the lack of coordination among actors limits the capability of the grid to enable the mentioned trends. In this paper we tackle this problem by proposing an architecture that enables households to autonomously exchange energy blocks and flexibility services with neighbors, operators and market actors. The solution is based on a blockchain transactive platform. We focus on a market application, where households can trade energy with their neighbors, aimed to locally balancing renewable energy production. We propose a market mechanism and dynamic transport prices that provide an incentive for households to locally manage energy resources in a way that responds to both prosumer and operator needs. We evaluate the impact of such markets through comprehensive simulations using power flow analysis and realistic load profiles, providing valuable insight for the design of appropriate mechanisms and incentives.

\end{abstract}

\section{Introduction}
Present electricity distribution grids were not designed for facing massive deployment of renewable energy sources (RES) on residential premises, nor for the structural changes in load induced by the expected fast growth of the electric car market.
 
Two key consequences are, that the opportunities rose by local RES as a major component of the energy transition cannot be fully leveraged, and that, with no architectural paradigm changes towards exploiting demand side flexibility\footnote{Ability to modify the electricity demand/supply profile.}, the required investments on infrastructure reinforcement may be extremely high \cite{BURGER}.
 
In this paper we propose a novel cost effective architecture that, on the one hand, enables households to autonomously exchange energy blocks with their neighbors in order to optimize their electricity bill, and on the other hand, enables the Distribution System Operator (DSO) to induce a global behavior that supports the stability of the network, by favoring the balance of the excess of renewable energy with flexible demand at the neighborhood level. 
 
The architecture relies on local energy markets, for which we propose an auction mechanism in charge of the efficient matching of households offers to buy and sell renewable energy. We exploit dynamic transport fee rebates for creating incentives for households to adapt their demand of electricity to the availability of local renewable energy production. Proposed mechanisms have the potential to alleviate congestion, reduce losses and consequently augment RES hosting capacity.
 
To implement the market we propose a blockchain-based transactive platform designed to enable the trustworthy execution of transactions among actors, thanks to increased security and transparency with respect to a centralized auctioneer, leading to simplified and automated system audits\cite{Monax}. 

We developed a proof of concept realization of the proposed market mechanism, which allows us to evaluate the impact of local renewable energy markets on distribution grid quality of supply through comprehensive simulations using power flow analysis and realistic load profiles.
 
Previous work on the exchange of energy among households has mainly focused on continuous double auctions\footnote{Buying or selling offers are matched as they arrive in the order of arrival.}
\cite{CDAscalability,intraCDA}, which are hard to implement over a distributed platform that lacks fine-grained transaction arrival ordering, while call markets in which orders are matched periodically\cite{HybridImmune,multi-round}, are better adapted and can also be more efficient\cite{Parsons}.
Most articles taking into account the impact of such markets on the distribution grid rely on locational marginal pricing\cite{powermatcher,micromarket} 
or message passing\cite{picard}, which requires specific knowledge on network topology and may penalize households depending on their location.
There is a recent trend towards local/micro markets for the exchange of energy among households over distributed platforms. Our work is the first to propose a thorough assessment of the impact of such markets on distribution grid quality of supply. 

Current blockchain-based approaches\footnote{\url{http://brooklynmicrogrid.com}, \url{https://powerledger.io}, \url{http://solarcoin.org}} are still in its infancy. They are not taking existing distribution grid needs into account, or they are trading renewable certificates rather than balancing renewable energy locally\cite{nrgcoin}. 

The main contributions of our work are the following:

Section \ref{sec:System} - Local renewable energy balancing markets: We propose the transaction of energy among households through a market aimed to locally balancing the excess of renewable energy production with flexible demand in the context of future low voltage distribution systems.

Section \ref{sec:implement} - Transactive platform implementation: We implement the market over a blockchain-based transactive platform that provides increased security, transparency and auditability with respect to centralized approaches.

Section \ref{sec:marketMech} - Market mechanism and price incentives: We propose an auction mechanism that is asymptotically efficient, truthful, weakly budget-balanced, and fair in the attribution of quantities. Additionally, we use simple transport fee rebates to incentive houses to locally control their flexible demand in a way that benefits both prosumer\footnote{Consumer evolution towards pro-active participation on grid activities.} and DSO.

Section \ref{sec:simul} - Distribution Grid impact assessment: We thoroughly assess the impact of markets on distribution grid quality of supply. For this, we consider a broad set of metrics (Section \ref{sec:metrics}) and we provide a simple but comprehensive model for the household optimization problem (Section \ref{sec:optim}). 

\section{System description}
\label{sec:System}
\begin{figure}[tb!]
\centering
\includegraphics[width=2.5in]{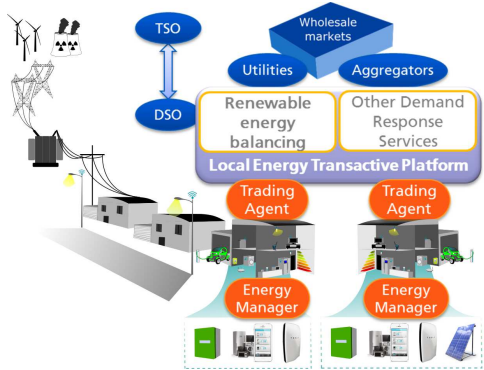}
\caption{Hierarchical architecture for distribution grid energy management.}
\label{fig_hierarchitecture}
\end{figure}

The low voltage distribution grid under study consists of one Medium Voltage/Low Voltage (MV/LV) transformer and one or more feeders to which residential prosumers are connected\footnote{This can also be seen as a Residential Virtual Distribution Grid\cite{Horta}.}, as shown in Figure \ref{fig_hierarchitecture}, in a context with high deployment of Photovoltaic Panels (PV) and rechargeable stationary batteries. We consider houses with a PV panel and a battery, houses with only a battery and houses without distributed energy resources\footnote{Flexible loads, controllable generation and storage resources.} (DER); all three types of houses count on smart meters. Batteries represent the main source of flexibility, as we consider inelastic demand, so that there is no impact on comfort and minimal user interaction is required. We also assume the flows from the PV and to/from the battery can be controlled by means of smart inverters
. This enables household to control the destination of the excess of renewable energy as well as the source of the energy that satisfies load demands.

Households interact with each other and with DSO, suppliers and service providers on the low voltage distribution grid through a local energy transactive platform to coordinate the access to DER. These actors are represented by agents autonomously transacting energy resources with each other in order to meet their economic goals, while collaborating with local and global grid infrastructure operation.
Households count on an energy management function (HEMS) for optimizing the allocation of DER depending on prosumer policies and economic incentives, and a Trading Agent function with the capability to exchange resources over the transactive platform.

\subsection{Local renewable energy balancing markets}
\label{subsec:localMarket}
This work focuses on renewable energy balancing markets at the neighborhood level, where households can exchange their excess/lack of energy. Based on the forecasted gap between consumption and production, the HEMS defines the energy available to trade in order to minimize the electricity bill. Then, the offers received from households' Trading Agents are matched by an auction algorithm that defines the accepted quantities to be exchanged and corresponding prices.

In terms of time scales, we focus on intra-day exchanges for energy blocks corresponding to 10 minutes periods, which would enable to timely update forecasts. Each market period will achieve finality at least one hour ahead of the time for which the resources are being allocated. This provides time for market actors to adapt their wholesale market offers, as the energy not balanced locally is provided/absorbed by utilities.

In order to evaluate the potential of proposed markets to augment variable RES hosting capacity, we need to consider appropriate quality metrics (Section \ref{sec:metrics}). In contrast with data networks, electricity is not blocked or delayed by lack of capacity, as distribution grids tend to have a relatively low utilization factor; while supporting high maximum flows increases losses, reduces transformer and cable lifespan and can cause voltage issues. For this reason, rather than focusing on capacity, we consider aggregated and maximum flows through the transformer, the peak to average ratio (PAR), and respective losses; while for the lines we also analyze losses as well as voltage deviations and network imbalance.

\subsection{Prices and auto-consumption}
\label{subsec:price}
In coherence with a context of massive deployment of RES, we assume that parity has been achieved, i.e.: electricity prices are higher than the Levelized Cost of Electricity (LCOE) produced by the PV. We consider the Feed In Tariff (FIT)\footnote{Regulated price that utilities must pay to prosumers for the renewable energy they inject into the grid.} to be lower than electricity prices, leaving a gap that can be exploited to establish a market for the exchange of energy among households. In this context, we assume houses will always prefer to auto-consume the energy available from their PV. Then, in cases of excess of production (resp. demand), households would have the following options for the surplus (resp. lack) of energy: they can charge (resp. discharge) their battery, inject on the grid for the FIT (resp. buy it from the utility) or sell (resp. buy) on the local renewable market.

Considering low FIT has two main motivations, first the reduction on LCOE from PV and the related increase in their deployment, and second, the need to incentive injection of renewable energy only when provides value to the grid. We propose that the gap between current FIT\footnote{Current FIT values in France are around 3 times the price of electricity, regardless of the suitability of the injection for the distribution grid.} and the FIT assumed by this work, could be used as an extra incentive for locally balancing renewable energy. Thus, only the energy that is certified to be balanced locally (thanks to the market) gets the extra feed in rewards; simultaneously motivating agents to enforce their contracted quantities through the local market. Furthermore, compared to current FIT, our approach not only benefits those who invest in PV, but also those who provide the flexibility necessary to host RES. Nevertheless, this incentive is not enough to take into account the needs of the DSO, as dynamic incentives need to be provided for households to charge their batteries when there is local renewable energy available. Such incentives will be proposed in Section \ref{subsec:priceIncent}.
 
\section{Transactive platform implementation}
\label{sec:implement}
Creating distributed transactive applications for the secure and transparent exchange of assets over a shared infrastructure formed by a network of competing parties has recently become feasible thanks to blockchain technology. 

The network is composed of validator nodes, for instance implemented by operators and service providers, in charge of forwarding and processing transactions into blocks, and validating the addition of new transaction blocks on a shared ledger. This shared ledger works like a replicated state machine, in which a new block represents a distributed agreement on its next state. The ledger can host code in the form of a smart contract\cite{Monax}
, which is executed each time a transaction gets addressed to it. This behavior is replicated across validator nodes, and allows the system to process transactions securely and reliably even in the presence of malicious nodes (Byzantine fault tolerance).

We use blockchain technology to build a chain of ownership over energy resources, i.e.: energy/flexibility units, so these can be traded in advance, in exchange of money in the form of a token payment. The parties involved in a transaction also receive the extra FIT reward as described in Section \ref{subsec:price}, as long as the enforcement of the corresponding electricity flows is validated by DSO\footnote{Such verification could be automatically done by smart meters.}. Such rewards are made through a token we call ECOin, representing a certificate of green energy traded locally, which can have many applications related to energy efficiency or CO2 emission reductions. 

\begin{figure}[tb!]
\centering
\includegraphics[width=3.0in]{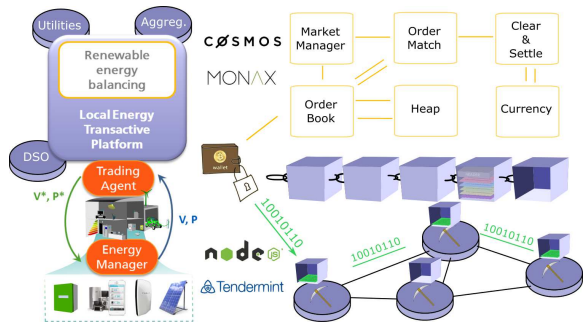}
\caption{Blockchain-based transactive platform and market implementation.}
\label{fig_testgrid}
\end{figure}

We used networks of Smart Contracts to implement each aspect of the market as exemplified on Figure \ref{fig_testgrid}, including registries (Renewable energy registry), matching offers (through implementation of auction mechanisms), clearing and settlement of transactions (verification of flows and attribution of ECOins). The Monax
platform was used to create the network of validator nodes, which achieve distributed agreement thanks to the Tendermint 
Byzantine Fault Tolerance algorithm\cite{Monax}. Monax provides also the tools to deploy smart contracts and Node.js libraries that were used for the Trading Agents to interact with the network of validators.

In addition to the local balancing markets, the distributed nature of blockchain-based markets will enable any aggregator or supplier to offer complementary demand response services. For instance, the deployment of bi-level optimization mechanisms considering competing aggregators, such as in \cite{aggregators}, could be done through smart contracts enabling prosumers to transact their flexibility without the need of having a previously signed legal contract. Furthermore, the shared transaction history could be used as a source of reputation of the different service providers as well as of the prosumers.

\section{Metrics for evaluating market impact}
\label{sec:metrics}
In this section we describe the quality metrics considered for the evaluation of the proposed renewable energy balancing market. We consider periods of 24 hours, slotted in timeslots of 10 minutes duration, $t \in \mathcal{T} = {1,...,T}$, where $T = 144$ is the total amount of timeslots. Let $P_{in,t}$ be the power flowing through the transformer from MV towards LV at timeslot $t$, expressed as $P_{in,t} = \sum_{ph}\max(P^{P}_{ph,t},0)$, where $P^{P}_{ph,t}$ is the power on the primary windings for phase $ph$, and $P_{out,t} = \sum_{ph}\max(-P^{S}_{ph,t},0)$ the flow from LV to MV, where $P^{S}_{ph,t}$ is the power on the secondary windings. The total power traversing the transformer is given by ${P_{abs,t} = P_{in,t} + P_{out,t}}$. Given the flows through the transformer we define corresponding losses as follows:
\begin{subequations}
\begin{align}
& \mbox{$Tr_{loss} = \sum_t(P_{abs,t}(C_L + C_C))$}\\
& \mbox{$L_{loss} = \sum_{l,ph,t}R_{l,ph}I^2_{l,ph,t}$}
\end{align}
\end{subequations}
where $C_L$ and $C_C$ are the coefficients representing leakage inductance losses and copper losses due to currents flowing through transformer windings, respectively. On the line losses, $R_{l,ph}$ is the resistance of the phase $ph$ of section $l$ of the distribution lines, and $I_{l,ph,t}$ the corresponding current.

With respect to the peak of power through the transformer:
\begin{subequations}
\label{eq:congestion}
\begin{align}
& \mbox{$P_{max} = \max(P_{abs,t})$ }\\
& \mbox{$PAR = T\frac{P_{max}}{\sum_t{P_{abs,t}}}$}
\end{align}
\end{subequations}

RES hosting capacity is affected also by voltage issues such as maximal deviations of voltage from its nominal value as well as network imbalance. The nominal value $V_U$ and the voltage deviation $V^{ph}_{delta}$ for each phase $ph$ are defined as follows, with $V^{ph}_{L2N}$ being the line to neutral voltage:
\begin{subequations}
\label{eq:DeltaV}
\begin{align}
& \mbox{$V_U = \frac{410}{\sqrt{3}}$}\\
& \mbox{$V^{ph}_{delta} = 100\frac{|V^{ph}_{L2N}| - V_U}{V_U}$}\\
& \mbox{$VUF = 100\frac{|V_2|}{|V_1|}$}
\end{align}
\end{subequations} 

Where $V_1$ and $V_2$ are the positive and negative voltage as per the symmetrical components method for unbalanced power systems and $VUF$ is the Voltage Unbalance Factor.

\section{Market mechanism}
\label{sec:marketMech}
The role of a market mechanism is to determine the final quantities and prices traded by each participant, by matching their orders. The exchange of energy blocks among trading agents will be carried out in a call market, through a Multi-unit Double Auction 
(MDA) mechanism\cite{Parsons}, where agents submit a bid or ask order, which are respectively the maximum, or minimum, price the agent is willing to pay to buy, or to accept to sell, the corresponding units of energy on certain timeslot. We propose a discrete-time MDA mechanism based on the algorithm proposed in\cite{MDA}, a variation of which has already been used for energy allocation markets in the context of electric vehicle charging\cite{Alonso} and complies with certain interesting characteristics: 
\begin{itemize}
\item Strategy-proof with respect to reservation price - Each agent's optimal strategy is to reveal truthful information regardless of other agents actions.
\item Weakly budget-balanced - All payments between buyers and sellers sum to a positive value, which is used to reward the nodes that process the transactions. 
\item Individually rational - The market encourages participation by ensuring non-negative profits.
\item Asymptotically efficient - The market becomes more efficient with the increasing amount of participants, maximizing total profit obtained by all participants.
\end{itemize}

The market will work on daily periods with one round to decide for each 10 minute timeslot, during which the offers are considered static. For any given round, each buyer $i$ submits a bid for $v^b_i$ units at a reservation price of $p^b_i$ and each seller $j$ an ask for $v^a_j$ units at $p^a_j$. The mechanism first generates the demand/supply curve, which sorts bids and asks in decreasing and increasing order of their reservation price, respectively. Then a buyer $B$ and a seller $S$ are identified at the critical point where the aggregate demand and supply meet. In order to make the mechanism strategy proof\cite{MDA} the corresponding houses do not participate in the trade, as well as the offers with index over $B$ and $S$. As a consequence of eliminating the critical offers, there are two cases for attributing quantities. 

Let the aggregated volumes of participating bids and asks be $V_B = \sum^{B-1}_1v^b_i$ and $V_S = \sum^{S-1}_1v^a_j$ respectively, then if $V_B \geq V_S$, all sellers with index $j < S$ trade their volume $v^a_j$ at price $p^a_S$, while all buyers with index $i < B$ trade at price $p^b_B$ a volume equal to $v^b_i\frac{V_S}{V_B}$; else if $V_S \geq V_B$, then all buyers with index $i < B$ trade their volume $v^b_i$ at price $p^b_B$, while all sellers with index $j < S$ will trade at price $p^a_S$ a volume equal to $v^a_j\frac{V_B}{V_S}$. This is as efficient as in \cite{MDA}, but is fairer on the individual attributions, from which the burden $|V_B - V_S|$ is subtracted proportionally to the offer rather than uniformly.

\subsection*{Price incentives}
\label{subsec:priceIncent}
In addition to the extra FIT reward proposed in Section \ref{subsec:price}\footnote{The analysis and simulations do not consider the extra FIT incentive.}, a dynamic incentive needs to be considered to synchronize battery charging with locally available renewable power. We propose the form of the incentive to be a rebate on the distribution/transport fee for those who buy electricity from the local market, as every $wh$ exchanged through the market is assumed not to traverse the transformer. We propose the value to be proportional to the day-ahead forecast of the average renewable energy production of a house on the neighborhood. Let $r^m$ be the maximum rebate considered, $p^u = {p^u_1,...,p^u_T}$ to be the utility prices for the day, and $\hat{g}^n = {\hat{g}^n_1,...,\hat{g}^n_T}$ be the forecast for the average renewable production, where the values are normalized so that $\hat{g}^n_t \leq 1 \; \forall t$. Then the minimum reservation price for each house at timeslot $t$ would be $p^r_t = p^u_t - \hat{g}^n_tr^m$ rather than $p^u_t$, which means that houses can potentially win more by making bids to charge their batteries when there is more renewable energy available.

\section{Households problem definition}
\label{sec:optim}
We consider a hierarchical architecture where the control on the distribution grid level, based on market mechanisms as explained above, can be designed independently from the control mechanisms inside the households, which is a benefit with respect to more tightly coupled models such as bi-level optimization\cite{aggregators}. For illustrating the feasibility of our approach\footnote{For a business model analysis (out of the scope of this paper) see \cite{BURGER}.}, a simple model of the problems that need to be solved by the HEMS is described. For the purpose of this article we consider a deterministic scenario in which consumption and production curves are known in advance\footnote{We do not address any particular forecasting mechanism/technology.}. This simplifying assumption relies on the capacity of batteries to absorb forecasting errors, particularly if forecasts were zero mean biased, and on the possibility of establishing similar market mechanisms closer to real time. 

\subsection{Determine offer quantities}
Let $h \in \mathcal{H} = \{0,...,23\}$ denote the next hour for which the local renewable energy market will accept offers. The HEMS needs to find optimal offer quantities $\hat{v}^o_{t_{0}+1},...,\hat{v}^o_{t_{0}+6}$, where $t_0 = 6h$ is the base timeslot of the hour $h$ for $6$ timeslots per hour. The energy units to offer on the market depend on the optimal use of the battery. The amount of energy stored on the battery at the end of the timeslot $t$ is represented by $e_{t}$ and its evolution in time is given by ${e_{t} = e_{t-1} - s_{t}}$, where $s_{t}$ represents the amount of energy charged or discharged to/from the battery during timeslot $t$. We require batteries to regain the initial state of charge at the end of the period $e_{T} = e_{0}$, so as to each study be independent, and also not to exceed one entire charge/discharge cycle of maximum 80\% depth of discharge per day, to extend battery life.

To satisfy such battery constraints, the HEMS needs to optimize over a rolling window formed by timeslots $t \in \mathcal{T}_{h} = \{t_0+1,..,144\}$, with $\tau_{h} = |\mathcal{T}_h|$. Each household has a (perfect) forecast of consumption $\ell_{h,t}$ and production $g_{h,t}$ for the rest of the day, from which it determines the gap $d_{h,t} = \ell_{h,t} - g_{h,t}$ that needs to be obtained by charging/discharging the battery or buying/selling on the local market. Households have the incentive to offer on the local market all the energy that is not exchanged with the battery, as they pay at most their bid limit price $p^b_{h,t}$ when buying and receive at least their ask limit price $p^a_{h,t}$ when selling.

Then, each household will look at the rest of the day and determine the best way of using the battery in order to minimize the electricity bill, depending on their limit prices and assuming that all the offers will be accepted. For each hour $h$ the HEMS will solve the following linear program:  

\begin{subequations}
\begin{align}
&\arg\min_{x_h} a^T_hx_h &a_h \in \Re^{3\tau_h},&&x_h \in \Re^{3\tau_h}\\
&\textrm{s.t.} &&&\nonumber\\
&x_{1,t} + x_{2,t} + x_{3,t} = d_{h,t} &&& \forall \; t \in \mathcal{T}_{h} \\
&-e_{h0} \le -\sum\nolimits^{j=t}_{j=t_0+1} x_{3,j} &\le E - e_{h0}&& \forall \; t \in \mathcal{T}_{h}\\
&-C^u \le x_{1,t} \le 0 &&& \forall \; t \in \mathcal{T}_{h}\\
&0 \le x_{2,t} \le C^u &&& \forall \; t \in \mathcal{T}_{h}\\
&|x_{3,t}| \le C^s  &&& \forall \; t \in \mathcal{T}_{h}
\end{align}
\end{subequations}
where $e_{h0}$ is the accumulated energy on the battery up to hour $h$ and $C^u$ and $C^s$ are the maximum power that can be exchanged on one timeslot with the grid and with the battery respectively. Variables ${x_{h,t} = (x_{1,t},x_{2,t},x_{3,t})^T}$ and deterministic coefficients ${a_{h,t} =(a_{1,t},a_{2,t},a_{3,t})^T}$ are as follows: 

\begin{subequations}
\begin{align}
&x_{1,t} : \hat{v}^a_{h,t} &&\textrm{Energy to sell on local market}\\
&a_{1,t} = \left\{
\begin{array}{ll}
p^a_{h,t},\\
0,
\end{array}
\right.&&\begin{array}{ll}
\textrm{if } g_{h,t} > \ell_{h,t},&\\
\textrm{else,}& 
\end{array}\nonumber\\ 
&x_{2,t} : \hat{v}^b_{h,t} &&\textrm{Energy to buy on local market}\\
&a_{2,t} = p^b_{h,t} && \nonumber\\
&x_{3,t} : s_{h,t} &&\textrm{Energy flow with the battery}\\
&a_{3,t} = 0 &&\nonumber 
\end{align}
\end{subequations}

With the solution we obtain the vector of offers $\hat{v}^o_h = \hat{v}^a_h + \hat{v}^b_h$ from which only the first 6 values corresponding to the next hour will be sent to the trading agent.

\subsection{Determine final flows}
Once the market results for each timeslot of the next hour $h$ are obtained by the trading agent, the HEMS needs to re-run the optimization problem in order to obtain the desired set point for the battery, as well as the flows to exchange with the utility, i.e.: the energy that is not exchanged with the battery will be sold for the FIT or bought from the supplier. The structure of the optimization problem is the same, except that the finally traded quantities $v^o_h = v^a_h + v^b_h$ are taken as an input and define the new gap to be satisfied $d_{h,t} = \ell_{h,t} - g_{h,t} - v^o_{h,t}$, and the optimization variables and parameters are as follows: 

\begin{subequations}
\begin{align}
&x_{1,t} : g^{FIT}_{h,t} &&\textrm{Energy to sell for the FIT}\\
&a_{1,t} = \left\{
\begin{array}{ll}
FIT,\\
0,
\end{array}
\right.&&\begin{array}{ll}
\textrm{if } g_{h,t} > \ell_{h,t},&\\
\textrm{else,}& 
\end{array}\nonumber\\
&x_{2,t} : \ell^u_{h,t} &&\textrm{Energy to buy from supplier } u\\
&a_{2,t} = p^u_{h,t} &&\textrm{Price offered by supplier } u\nonumber\\
&x_{3,t} : s_{h,t} &&\textrm{Energy flow with the battery}\\
&a_{3,t} = 0 &&\nonumber
\end{align}
\end{subequations}

\section{Simulations and Results}
\label{sec:simul}
\subsection{Simulation tools, procedure and assumptions}
We use power flow analysis as the means to assess the impact of the exchange of energy among households on quality of supply over the distribution grid. We rely on the Distribution Network Simulation Platform (DisNetSimPl) developed by EDF R\&D. The platform provides an interface to several simulation tools, from which we use the open source Distribution System Simulator OpenDSS. For illustrative purposes we designed an electricity network model conformed by a 20 kV/410 V transformer of 160 kVA rated power, two feeders with aluminum power lines of 240 $mm^2$ and neutral of 95 $mm^2$, following general specifications for French distribution grids
\footnote{\url{http://www.enedis.fr/sites/default/files/Enedis-PRO-RES_43E.pdf}}\footnote{Network file available at \url{https://github.com/joluHo/P2P-energy-trading}}. The experiments were based on realistic synthetic consumption data obtained from the Multi-agent Simulator of Human Behavior SMACH\cite{amouroux2013simulating}. The load curves correspond to 6 winter days of consumption from 33 households of mixed profiles. We requested data for intervals of 10 minutes in order to be able to simulate the system we define in Section \ref{subsec:localMarket}.

The daily load served as an input to the individual household control mechanism; more precisely as the consumption forecast of each household. After each intraday decision taken and the corresponding market interactions, the flows to be exchanged with the battery and to be transacted with the neighbors are calculated. The final load curve exchanged with the grid for each household is used as an input for the DisNetSimPl simulator to analyze the impact of the market on the distribution grid.

For the simulations we considered $N = 33$ households, $8$ without flexibility (approx $N/4$), $8$ with battery only (approx $N/4$) and $17$ with PV and battery (approx $N/2$).
We consider ideal batteries (without losses) of sizes, 3, 6 and 9 kWh. We consider a common supplier offering a Time Of Use pricing with two levels: 15 c\euro{}/kWh from 12 am to 4 pm and 30 c\euro{}/kWh from 5 pm to 11 pm\footnote{The usual peak price at midday does not fit the context of this work.}. The reserve prices of houses are randomly obtained from a uniform distribution with values ranging from 0 to 30\% improvement over alternative prices from the utility, while the price rebate is below 2 c\euro{}/kWh.

\subsection{Results and discussion}
Results are presented as the percentage enhancement on the metrics introduced in Section \ref{sec:metrics} with respect to the scenario without market\footnote{Results are also applicable to centralized auctioneer implementations.}. The results are shown for 6 days (where the only change on the simulation are the base load curves on every household) and for several battery sizes. For each simulation batteries in all households have the same size.

\begin{figure*}
    \centering
    \begin{subfigure}[b]{0.4\textwidth}
        \includegraphics[width=\textwidth]{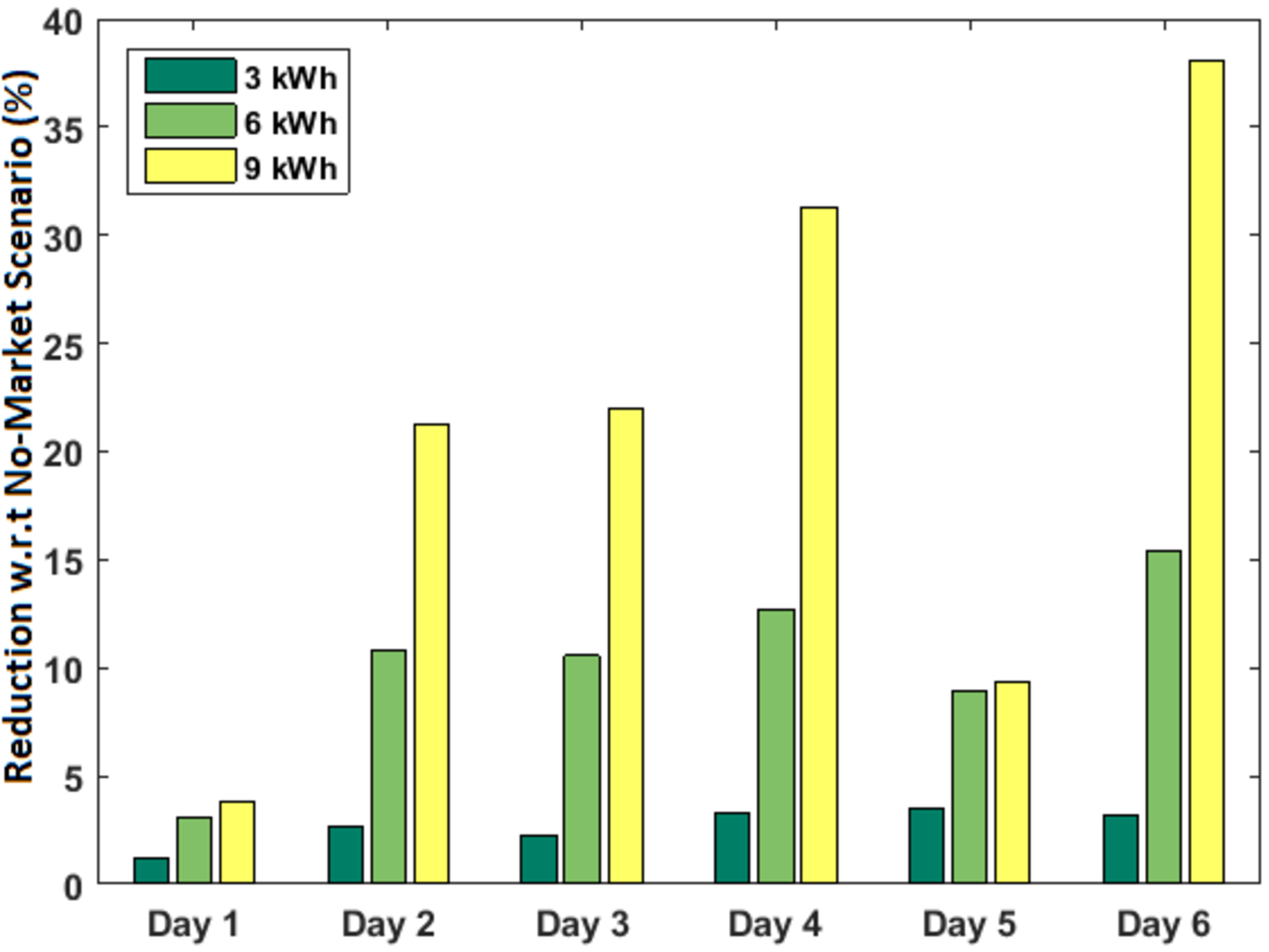}
		\caption{Energy out.}
        \label{fig_E_out}
    \end{subfigure}
    ~ 
    \begin{subfigure}[b]{0.4\textwidth}
        \includegraphics[width=\textwidth]{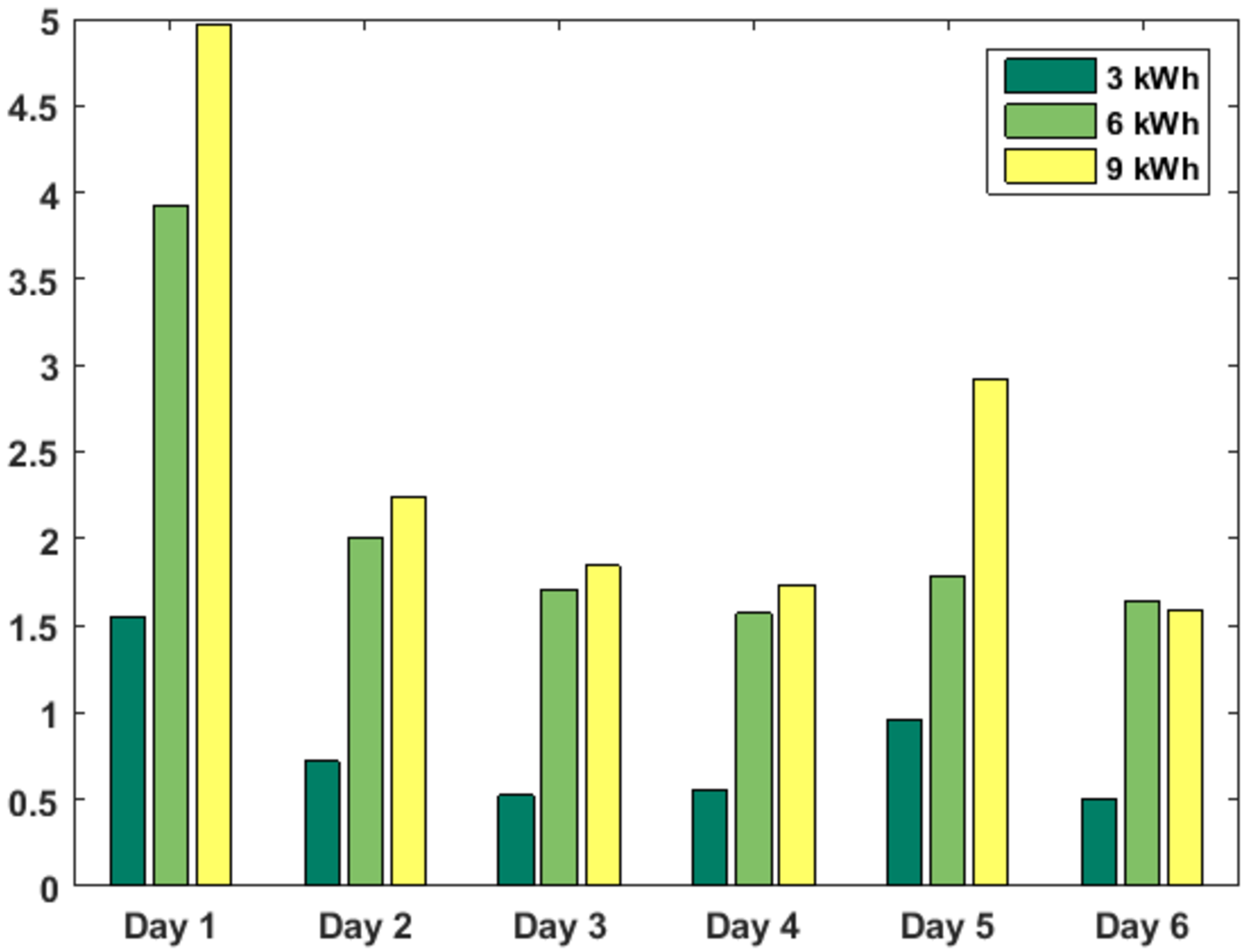}
		\caption{Energy in.}
        \label{fig_E_in}
    \end{subfigure}
    \caption{Energy through the transformer.}\label{fig:TransfLoadReduc}
\end{figure*}

\begin{figure*}
    \centering
    \begin{subfigure}[b]{0.4\textwidth}
        \includegraphics[width=\textwidth]{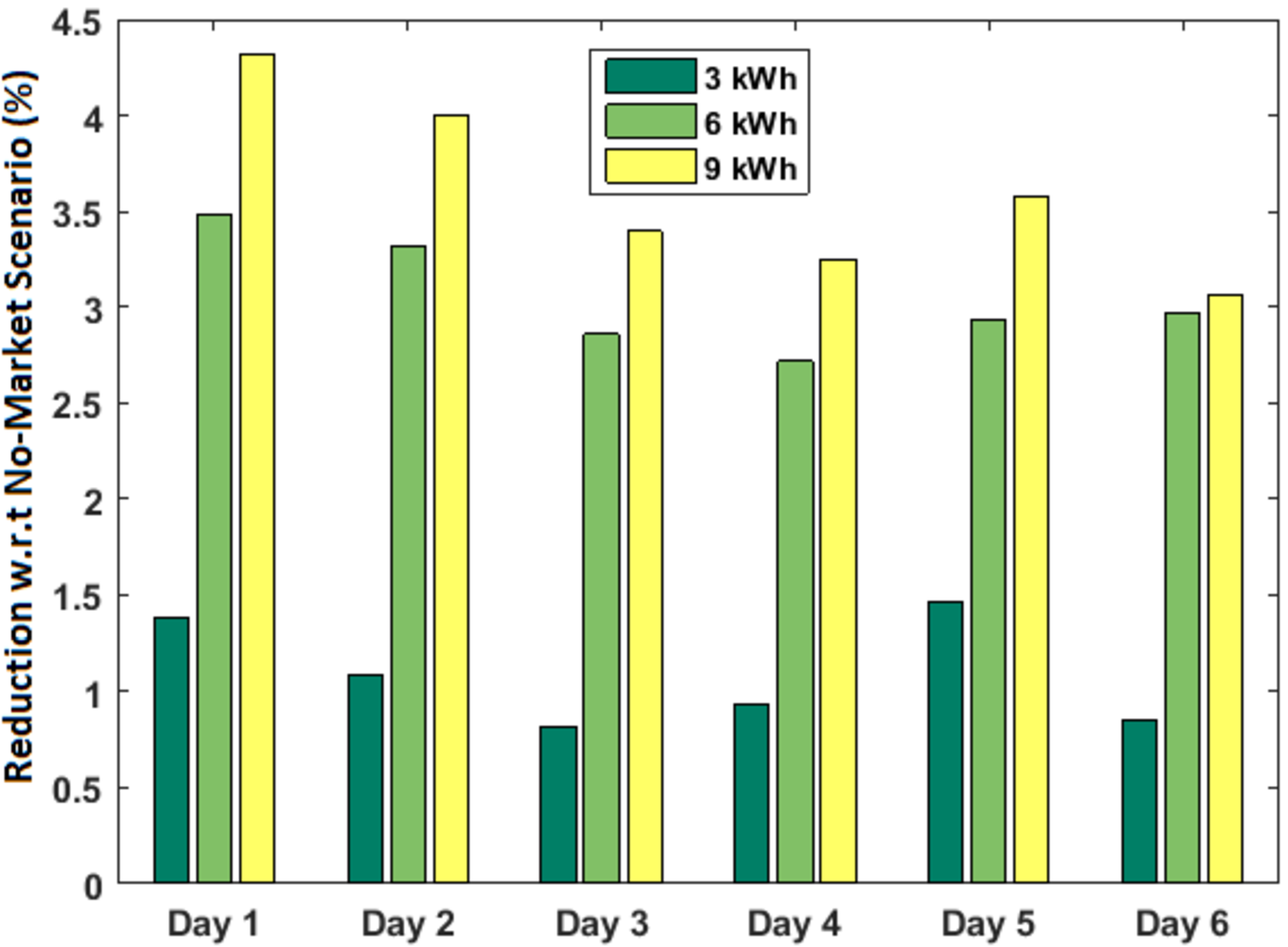}
		\caption{Transformer losses.}
        \label{fig_T_Loss}
    \end{subfigure}
    ~ 
    \begin{subfigure}[b]{0.4\textwidth}
        \includegraphics[width=\textwidth]{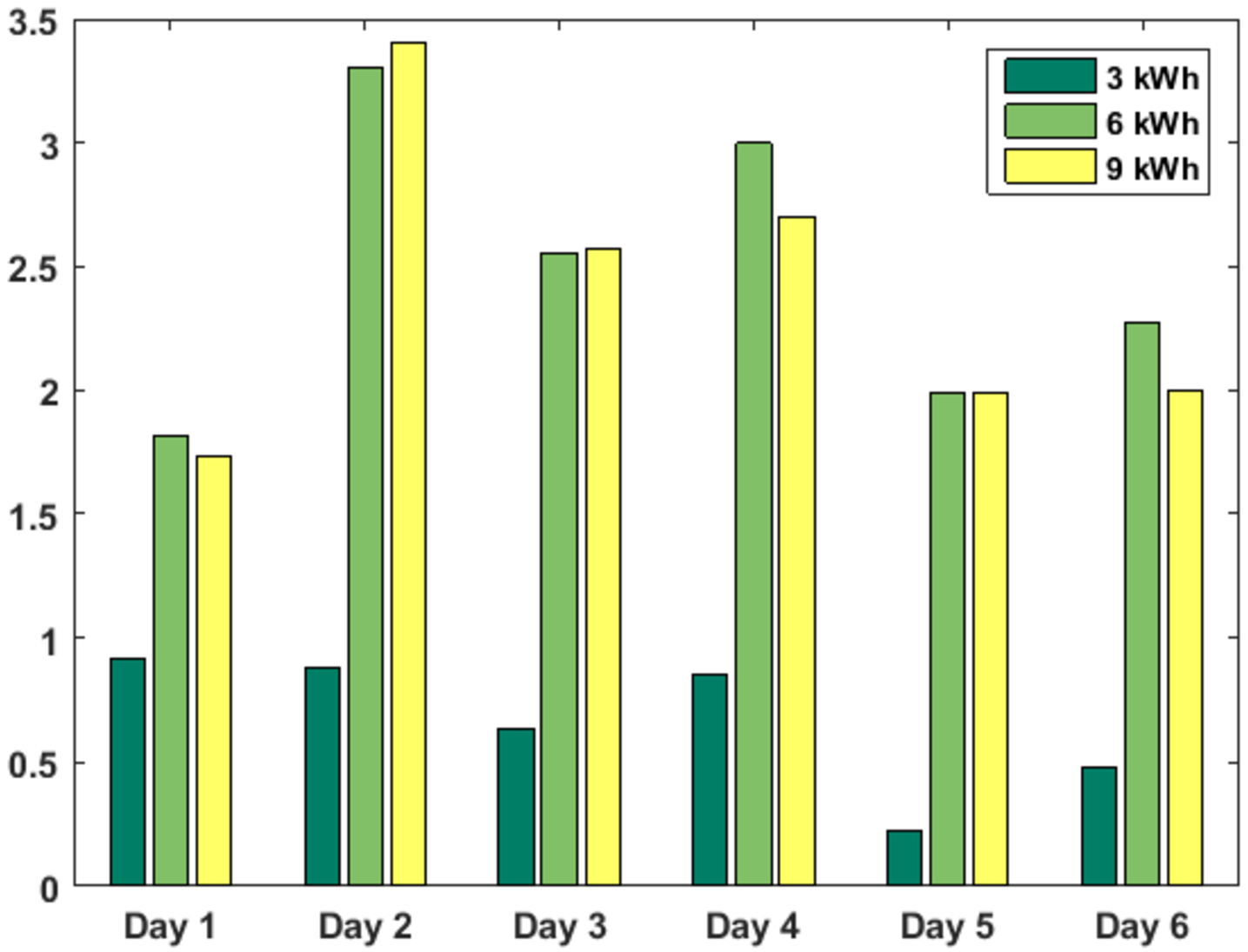}
		\caption{Line losses.}
        \label{fig_L_loss}
    \end{subfigure}
    \caption{Reduction of losses.}\label{fig:TotalLosses}
\end{figure*}

Figure \ref{fig_E_out} shows that a market enabling the exchange of energy among households coupled with appropriate incentives enables considerable reductions on the excess of renewable energy production towards Medium Voltage. The non-decreasing dependency with the size of the battery is coherent with the increase in flexibility available for households, and it is maintained despite variations on residential demand in different days, with a lower impact on days 1 and 5 due to a lower aggregated demand on the peak hours. For the understanding of the behavior of the system it is interesting to note that, as the size of batteries increase, the excess of energy in houses with PV is reduced and the offers to sell on the market as well. This can also be explained by the fact that when the units per bid is increased and the units per asks is decreased, the amount of bidders participating on the market can potentially be reduced, creating a negative impact on the efficiency of the strategy proof market mechanism. 

There is also a consistent reduction on the energy entering the transformer from the Medium Voltage, though the reduction is less considerable as it represents a small share of the total demand. The total reduction of flow through the transformer leads to a reduction on transformer losses close to 4\% for 6 and 9 kWh batteries, as we can see on Figure \ref{fig_T_Loss}. The total line losses are also reduced as shown in Figure \ref{fig_L_loss}.

\begin{figure*}
    \centering
    \begin{subfigure}[b]{0.4\textwidth}
        \includegraphics[width=\textwidth]{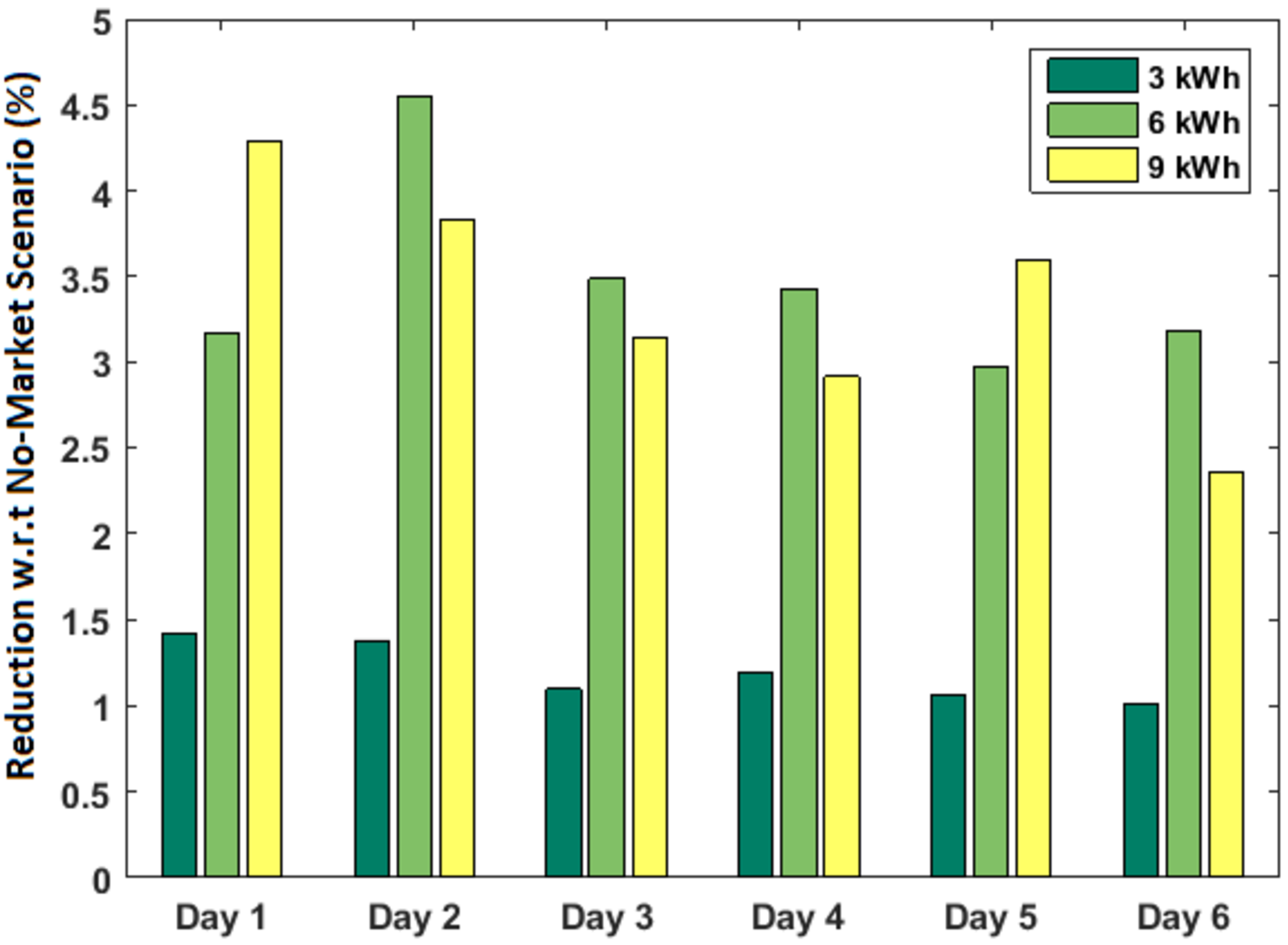}
		\caption{On phases.}
        \label{fig_Loss_ph}
    \end{subfigure}
    ~ 
    \begin{subfigure}[b]{0.4\textwidth}
        \includegraphics[width=\textwidth]{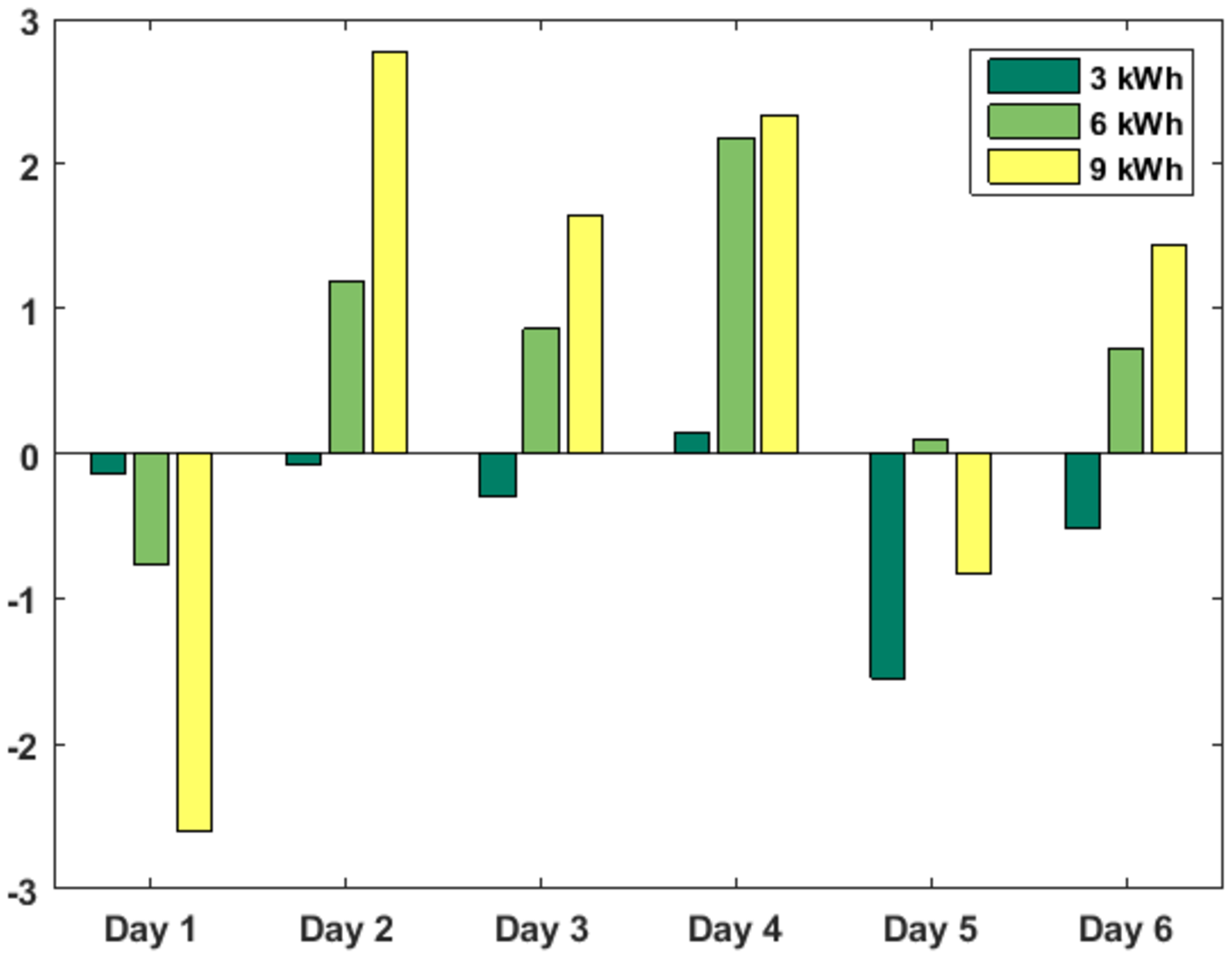}
		\caption{On neutral.}
        \label{fig_Loss_neutral}
    \end{subfigure}
    \caption{Reduction of line losses.}\label{fig:perLineLoss}
\end{figure*}

\begin{figure*}
    \centering
    \begin{subfigure}[b]{0.4\textwidth}
        \includegraphics[width=\textwidth]{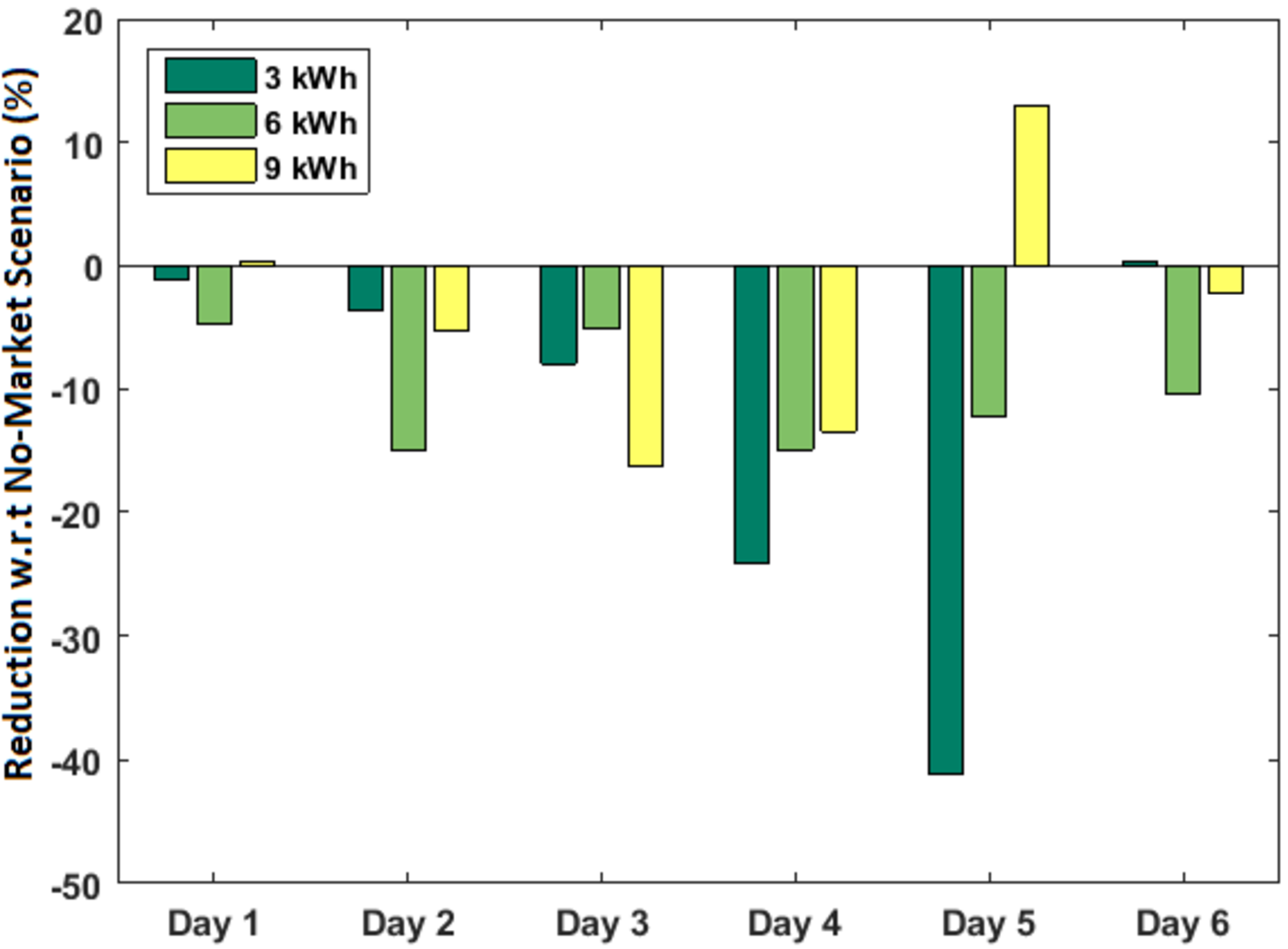}
		\caption{Maximum VUF.}
        \label{fig_VUF_max}
    \end{subfigure}
    ~ 
    \begin{subfigure}[b]{0.4\textwidth}
        \includegraphics[width=\textwidth]{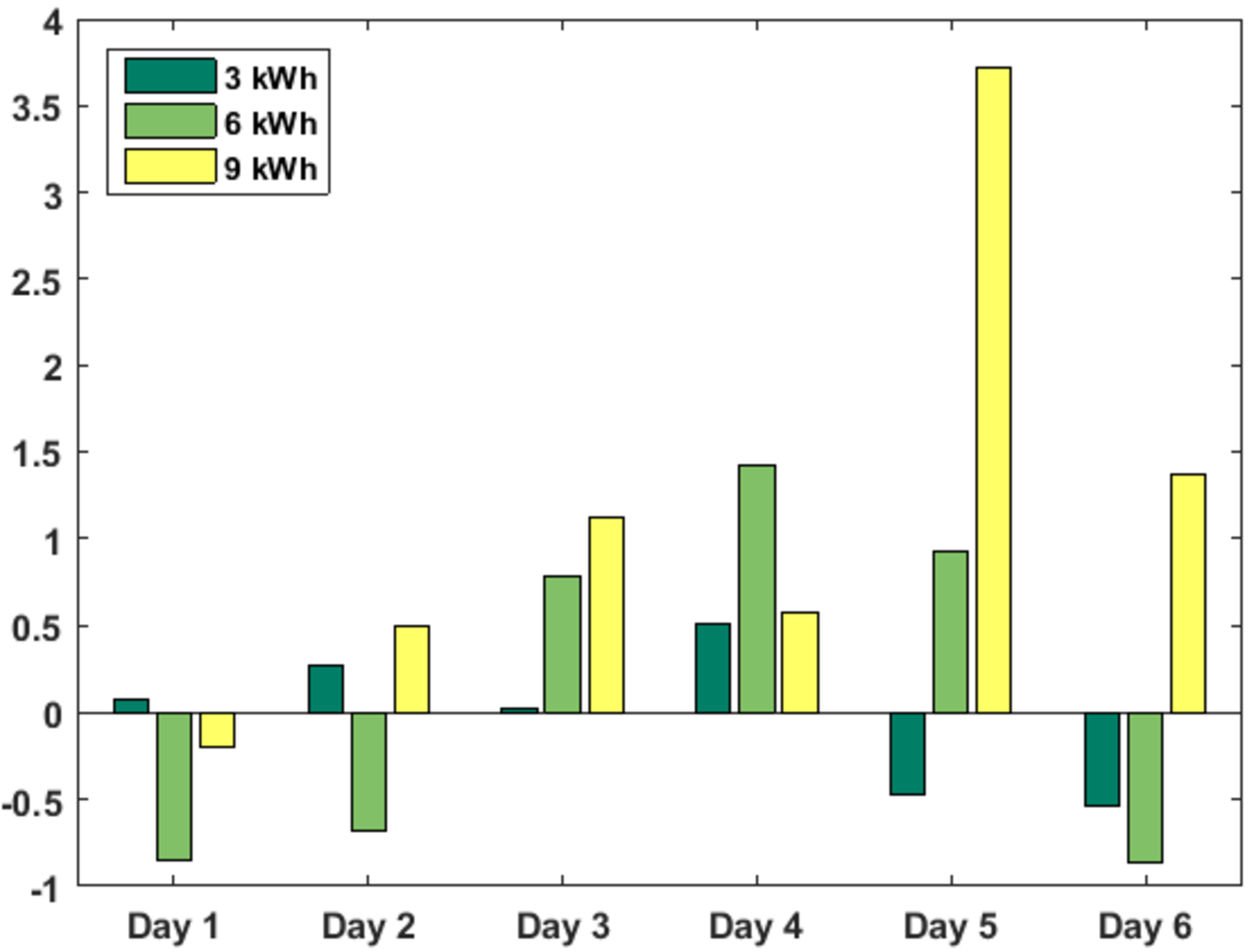}
		\caption{Mean VUF.}
        \label{fig_VUF_mean}
    \end{subfigure}
    \caption{Voltage Unbalance Factor.}\label{fig:VUF}
\end{figure*}

We can see, in Figure \ref{fig_Loss_ph}, that the market consistently achieves a reduction of losses in line phases. This is not the case for the neutral as shown in Figure \ref{fig_Loss_neutral}. The market mechanism does not take into account the phases when allocating the quantities to be traded, leading to production on one phase being balanced with consumption in another phase. These unequal flows on the phases unbalance the network provoking an increase on the neutral current and respective losses.

The unbalance effect of the market is reflected by an important increase on the maximum VUF as shown in Figure \ref{fig:VUF}.
From the overall results of our analysis it is understood that the only negative impact of such imbalance would be on triphasic loads connected to the network. We are working on new market designs to cope with this issue.

\begin{figure*}
    \centering
    \begin{subfigure}[b]{0.4\textwidth}
        \includegraphics[width=\textwidth]{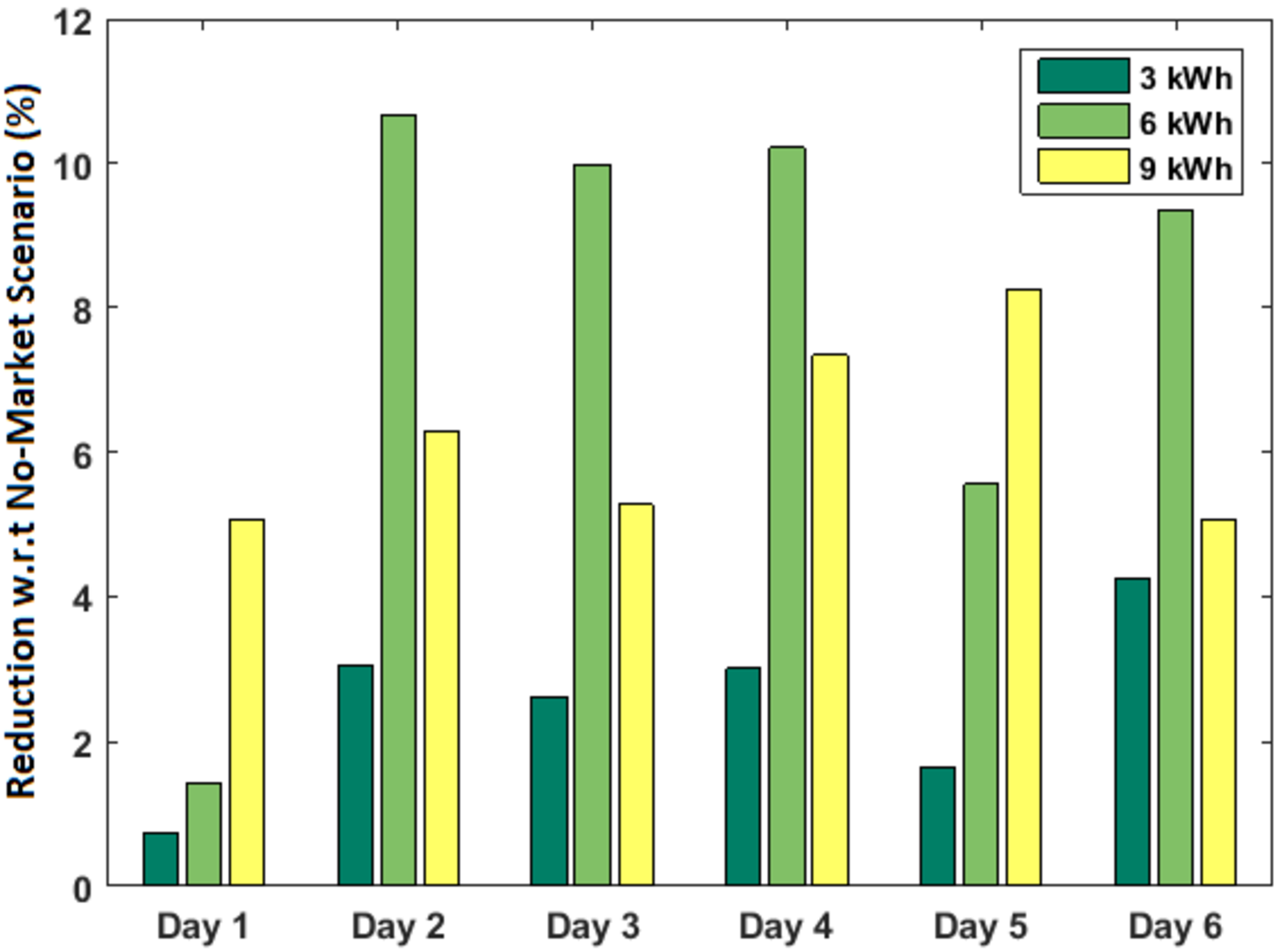}
		\caption{Maximum transformer load.}
        \label{fig_P_max}
    \end{subfigure}
    ~ 
    \begin{subfigure}[b]{0.4\textwidth}
        \includegraphics[width=\textwidth]{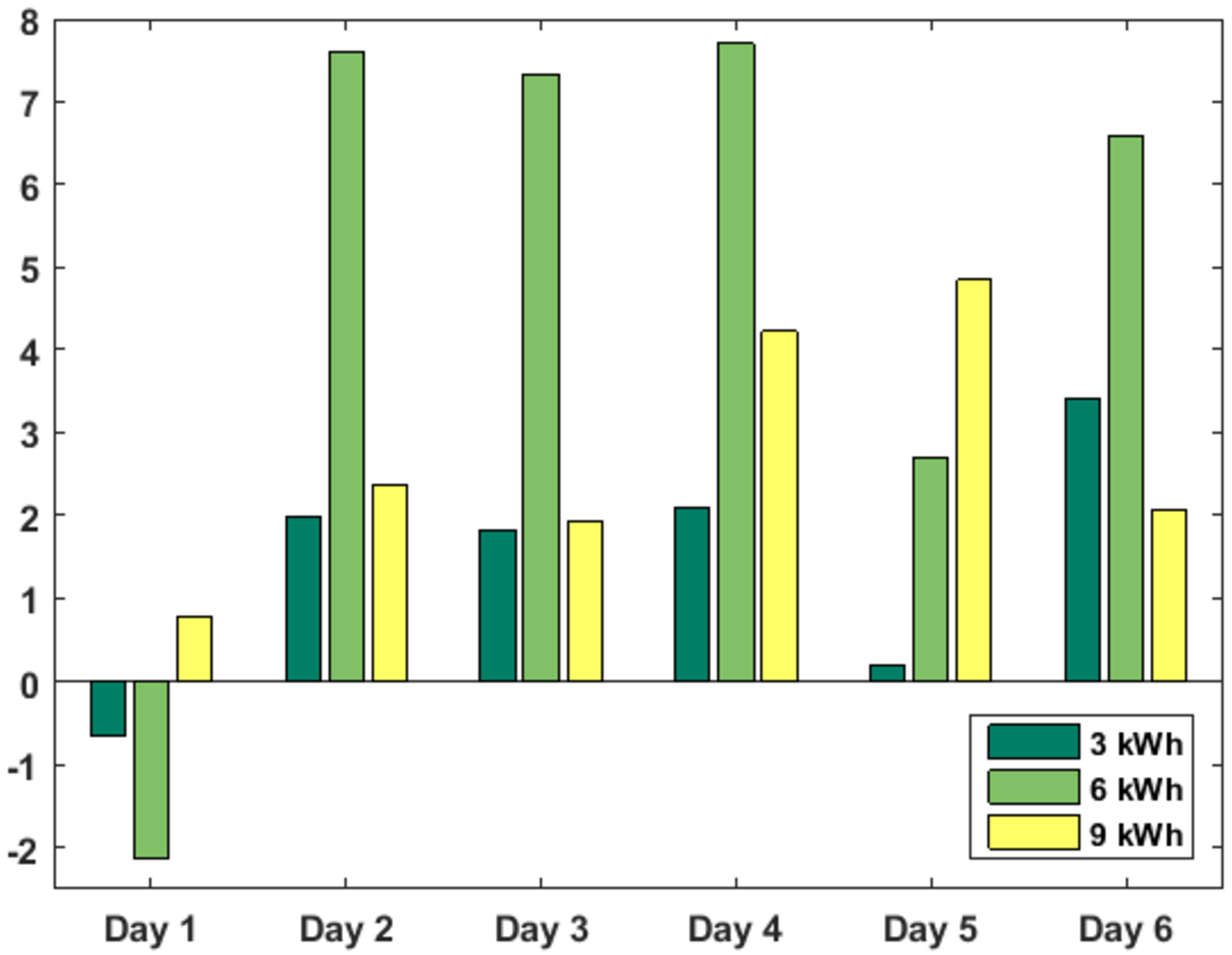}
		\caption{PAR.}
        \label{fig_P_mean}
    \end{subfigure}
    \caption{Grid congestion assessment.}\label{fig:congestion}
\end{figure*}

Figure \ref{fig_P_max} shows that the peak load supported by the transformer is consistently and considerably reduced. The PAR is in some cases increased, due to a more important reduction on the mean load, which is also positive for the grid.

\begin{figure*}
    \centering
    \begin{subfigure}[b]{0.4\textwidth}
        \includegraphics[width=\textwidth]{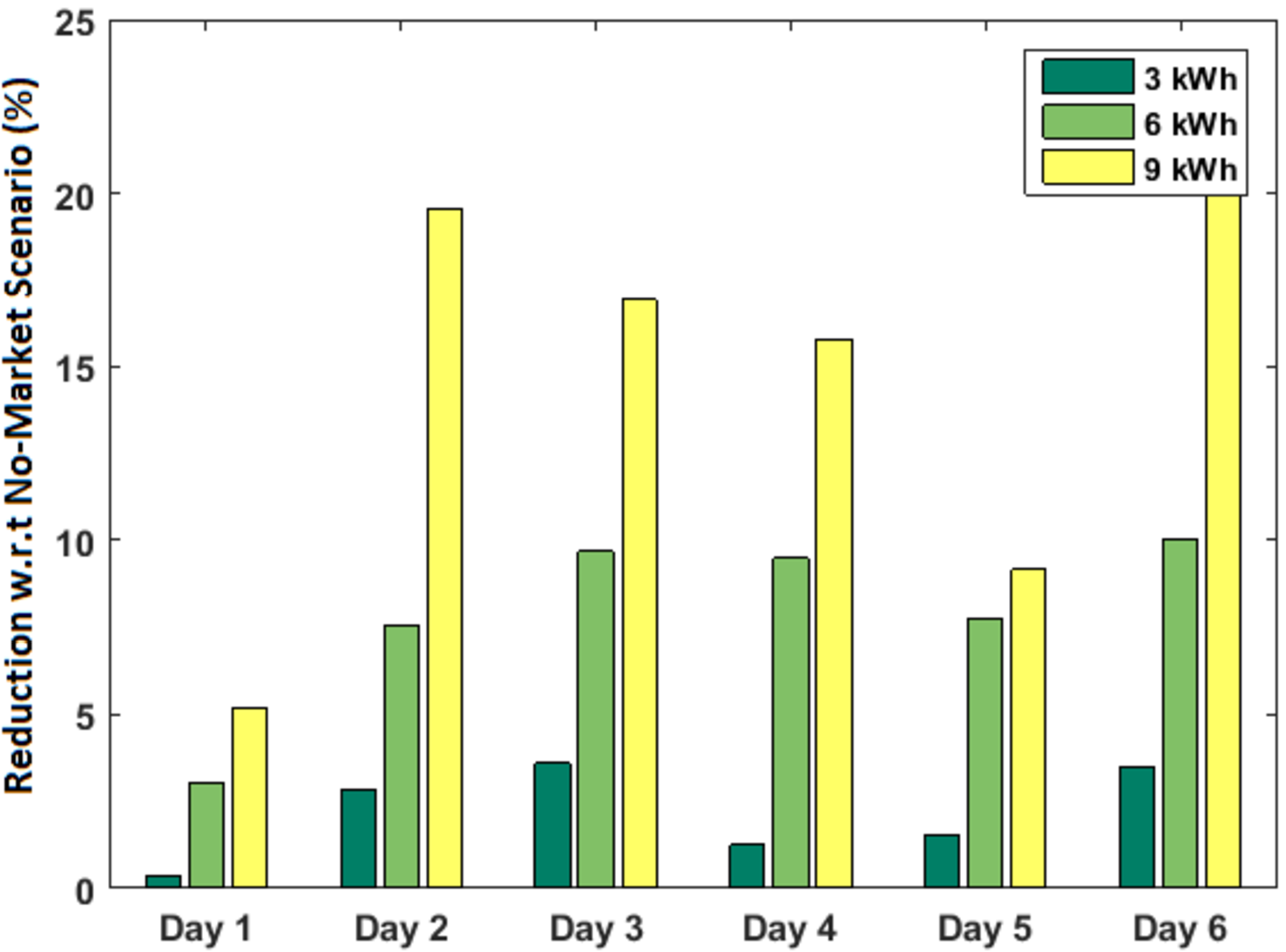}
		\caption{Maximum deviation.}
        \label{fig_V_max}
    \end{subfigure}
    ~ 
    \begin{subfigure}[b]{0.4\textwidth}
        \includegraphics[width=\textwidth]{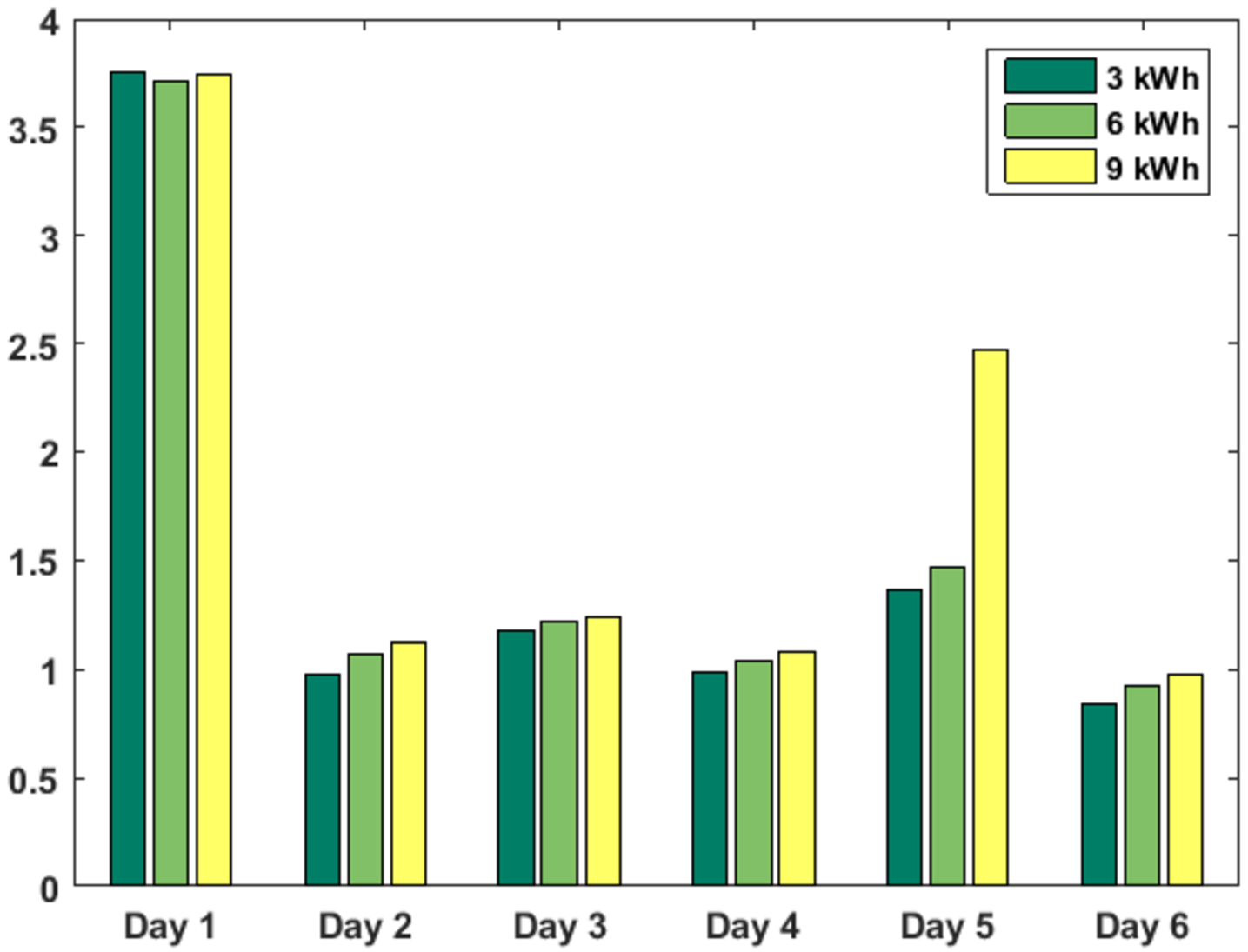}
		\caption{Mean deviation.}
        \label{fig_V_mean}
    \end{subfigure}
    \caption{Voltage profile deviation.}\label{fig:V_profile}
\end{figure*}

In current distribution networks voltage deviations depend on how loads are connected to the different phases and varies on time due to the load evolution of the different households. We observe that our system for some phases has positive impacts on voltage deviation, as shown in Figures \ref{fig_V_max} and \ref{fig_V_mean}, but simultaneously in other phase the impact can be negative. Nevertheless, in all cases the operation limits are respected. Control on individual line current limits or voltage deviations require additional mechanisms that are not considered in this paper. These are subject of future work, as they can function in parallel with the renewable energy market (as suggested at the end of Section \ref{sec:implement}), or at a time scale closer to real-time. 

\section{Conclusion}
Major ongoing evolutions in the electricity industry, like local deployment of RES and storage capacity, represent a key driver for the ``energy transition'' and related objectives. Nevertheless, the required transformations of the distribution grid are today delaying, and in some cases blocking, the possibility to leverage such opportunities. In this work we propose increasing RES hosting capacity by introducing autonomous markets for the exchange of energy among households aimed to locally balancing renewable energy production. The proposed system is based on auction mechanisms and dynamic transport fee rebates that serve as an incentive for households to control their energy resources in a way that benefits both prosumer and operator. The architecture relies on a blockchain-based transactive platform to enable the coordinated participation of any type of player, including DSOs, aggregators, and prosumers, thanks to its increased security, transparency and auditability with respect to centralized implementations.

To the best of our knowledge, our work is the first to propose a thorough assessment of the impact of the exchange of energy among households on distribution grid quality of supply, for which we use power flow simulations with realistic load data. The results show that the auction mechanism and price incentives proposed on this paper enable an increase of RES hosting capacity by reducing the maximum power flowing through the transformer and reducing losses on both the transformer and the lines. Moreover, the fact of locally balancing the excess of PV production reduces the possible negative impacts of massive PV deployment on the global grid. Additional benefits could be obtained if the mechanisms could reduce voltage deviation and network unbalance. An approach for coping with these issues could be enabling DSOs to actively provide incentives and constraints for the operation of DER. We are currently working on game theory models for establishing distributed mechanisms that enable households to determine optimal strategies for reducing their electricity bill, leading to a competitive aggregative equilibrium that respects, in particular, such distribution grid constraints. Moreover, our current work will extend simulations to scenarios with more houses and different penetration of resources, and will assess the impact of forecast uncertainty on the obtained results.

\bibliographystyle{hieeetr}
\bibliography{references}

\end{document}